\RequirePackage{lineno}
\documentclass[twocolumn,aps,prc,showpacs,superscriptaddress]{revtex4}
\usepackage{epsfig}
\def \lp {$\bar{\Lambda}/\bar{p}$}
\def \Lp {$(\bar{\Lambda}+\bar{\Sigma}^0)/\bar{p}$}
\def \LP {$(\bar{\Lambda}+\bar{\Sigma}^0+1.1\bar{\Sigma}^-)/\bar{p}$}
\def \lbar {\bar{\Lambda}}
\def \Lbar {\bar{\Lambda}+\bar{\Sigma}^0}
\def \pbar {\bar{p}}
\def \pbarp {\bar{p}p}
\def \lbarp {\bar{\Lambda}p}
\def \snn {\sqrt{s_{_{NN}}}}

\begin{document}
%\draft

%%%%%%%%%%%%%%%%%%%%%%%%%%%%%%%%%%%%%%%%%%%%%%%%%%%%%%%%%%%%%%%%%%%%%%%%%%%

\title{Effects of nuclear absorption on the \lp\ ratio in relativistic heavy ion collisions}
\author{Fuqiang Wang}
\affiliation{Department of Physics, Purdue University, 525 Northwestern Avenue, West Lafayette, Indiana 47907, USA}
\author{Marlene Nahrgang}
\author{Marcus Bleicher}
%\affiliation{Nuclear Science Division, Lawrence Berkeley National Laboratory, Berkeley, CA 94720, USA}
\affiliation{Frankfurt Institute for Advanced Studies (FIAS), Johann Wolfgang Goethe Universitaet, Frankfurt am Main, Germany}

\begin{abstract}
An enhanced \lp\ ratio in heavy-ion relative to $p+p$ collisions has been proposed as one of the signatures for the Quark-Gluon Plasma (QGP) formation. A significantly large \LP\ ratio of 3.5 has been observed in the mid-rapidity and low transverse momentum region in central Au+Au collisions at the nucleon-nucleon center-of-mass energy of $\snn=4.9$~GeV at the Alternating Gradient Synchrotron (AGS). 
%This is significantly larger than the value of 0.2 in $p+p$ collisions at the corresponding energy. 
This is an order of magnitude larger than the values in peripheral Au+Au collisions and $p+p$ collisions at the corresponding energy. 
By using the Ultra-relativistic Quantum Molecular Dynamics (UrQMD) transport model, we demonstrate that the observed large ratio can be explained by strong absorption of $\pbar$'s ($\sim$99.9\%) and $\lbar$'s ($\sim$99\%) in dense nuclear matter created in central collisions. We find within the model that the initial \lp\ ratio, mainly from string fragmentation, does not depend on the collision centrality, and is consistent with that observed in $p+p$ collisions. This suggests that the observed large \LP\ ratio at the AGS does not necessarily imply the formation of the QGP.
%Comparing the UrQMD results at SPS energy to preliminary data leads to a similar conclusion.
%In order to gauge possible large \lp\ ratios from higher energy heavy-ion collisions at RHIC due to QGP formation, we calculate the ratio for Au+Au collisions at RHIC energies keeping the same fragmentation functions as from $p+p$ collisions. We find that the absorption effect is much smaller ($<???$) at RHIC than the lower AGS energy.
We further study the excitation function of the ratio in UrQMD, which may help in the search and study of the QGP.
\end{abstract}


\maketitle

%%%%%%%%%%%%%%%%%%%%%%%%%%%%%%%%%%%%%%%%%%%%%%%%%%%%%%%%%%%%%%%%%%%%%%%%%%%

Strangeness production has been extensively studied in heavy-ion collisions because enhanced strangeness production may signal the formation of Quark-Gluon Plasma (QGP)~\cite{Muller,Rafelski,Koch}. This is due to the fact that the strangeness ($s\bar{s}$) production threshold is significantly lower in a QGP than in a hadronic gas in which a $s\bar{s}$ pair has to be produced by a pair of strange hadrons. Strangeness enhancement is often studied by the charged kaon production rate and the kaon to pion yield ratio ($K/\pi$). All experimental results showed an unambiguous enhancement in kaon production rate and $K/\pi$ ratio in heavy-ion collisions with respect to elementary $p+p$ collisions~\cite{E866kaon,E866kpi,NA49kaon,NA44kaon,STARkpi}. However, the enhancement results can be also explained by particle rescattering as implemented in many hadronic transport models~\cite{RQMDkpi}.

At AGS energies, the collision zone created in central heavy-ion collisions is baryon dense~\cite{BrichAGS,BrichSPS,WangBaryon}. In a QGP with high baryon density, production of light antiquarks ($\bar{u}, \bar{d}$) should be suppressed, hence the ratio of anti-lambda to antiproton yields (\lp) should exhibit a larger value of enhancement than does the $K/\pi$ ratio~\cite{Rafelski,Lee,Ko}. This is because \lp\ gets enhancement not only from enhanced strange antiquark production but also from the suppressed $\bar{u}$ production.

The E864 Collaboration at the AGS has deduced a \LP\ ratio in Au+Au collisions at the nucleon-nucleon center-of-mass (c.m.s.) energy of $\snn=4.9$~GeV. They deduced the ratio at mid-rapidity and almost zero transverse momentum ($p_T$) by contributing the discrepancy between their $\pbar$ measurement~\cite{E864PRL,E864PRC} and that from AGS/E878~\cite{E878PRL,E878PRC} entirely to the different acceptances of the two experiments for $\pbar$'s from $\lbar$ decays~\cite{E864PRL,E864PRC}. 
%The deduced \LP\ value, while being consistent with $p+p$ results in peripheral collisions at similar energies, has a strong dependence on the collision centrality. 
The deduced \LP\ value has a strong dependence on the collision centrality. 
In peripheral collisions, it is consistent with $p+p$ results ($\sim 0.2$) at similar energies~\cite{Blobel,Rossi}.
In most 10\% central collisions, the \LP\ ratio reaches a most probable value of 3.5, which is over an order of magnitude larger than those in peripheral collisions and in $p+p$ collisions. The E917 Collaboration has made direct measurements of $\pbar$ and $\Lbar$ yields at mid-rapidity and integrated over $p_T$ in central and peripheral collisions~\cite{E917}. The ratios of \Lp, given the large error bar, are consistent with E864. These results are intriguing because they may point to possible QGP formation at the AGS.

There are at least two physical origins for the large \LP\ ratio: (1) an enhanced \lp\ (and/or $\bar{\Sigma}^{0,-}/\bar{p}$) ratio at the initial production stage of the antibaryons, and (2) a strong absorption of $\pbar$'s and a less strong absorption of $\lbar$'s and $\bar{\Sigma}$'s in nuclear matter produced in heavy-ion collisions. An enhanced \lp\ ratio at the initial production stage would be evidence for QGP formation. In order to obtain the \lp\ ratio at the initial stage, one has to postulate from measurements at the final freeze-out stage including the nuclear absorption effects. To this end, we use the Ultra-relativistic Quantum Molecular Dynamics (UrQMD) model~\cite{UrQMD} to simulate Au+Au collisions at $\snn=4.9$~GeV and record the initial production abundances of antibaryons and high-mass antibaryon resonances as well as the final freeze-out abundances. We chose UrQMD because, as we discuss later, (1) it has been reasonably successful in describing many of the experimental results on hadron spectra, as well as the average baryon density and the baryon emitting source size which are the essential ingredients for nuclear absorption, and (2) it does not have a QGP state or mechanisms mimicking a QGP state. Our strategy is then to compare the final freeze-out \LP\ ratio to data and use the initial \lp\ information from UrQMD to conjecture about what the data might be telling us. We shall demonstrate that the observed large \LP\ is consistent with strong annihilation of $\lbar$'s and $\pbar$'s in the nuclear medium created in Au+Au collisions. Therefore, the data does not lead to a conclusion that QGP is formed.

%UrQMD is a transport model employing string degree of freedom. It has 
%important ingredient relavent to heavy-ion collisions: string excitation
%and fragmentation and hadronic binary collisions. 
%History of the collision is rigarously followed until final freeze-out 
%when particles cease to interact both inelastically and elastically.
%...more description of urqmd....

%As a tool for our investigation of heavy-ion reactions at RHIC the Ultra-relativistic Quantum Molecular Dynamics model (UrQMD 1.2) is applied \cite{UrQMD}. 

The UrQMD model has been applied successfully to explore heavy-ion reactions from AGS energies (E$_{\rm lab}=1-10$~AGeV) up to the full CERN-SPS energy (E$_{\rm lab}=200$~AGeV). This includes detailed studies of thermalization \cite{bravina}, particle abundancies and spectra \cite{Bass,Bleicher}, strangeness production \cite{Soff}, photonic and leptonic probes \cite{Spieles,Ernst} , $J/\Psi$'s \cite{jpsi}, and event-by-event fluctuations \cite{BleicherPLB,BleicherNPA}.

%Similar to the RQMD model \cite{rqmd}, 
UrQMD is a microscopic transport approach based on the covariant propagation of
constituent quarks and diquarks accompanied by mesonic and baryonic degrees of freedom. It simulates multiple interactions of ingoing and newly produced particles, the excitation and fragmentation of color strings, and the formation and decay of hadronic resonances. %At RHIC energies, the treatment of subhadronic degrees of freedom is of major importance. In the UrQMD model, these degrees of freedom enter via the introduction of a formation time for hadrons produced in the fragmentation of strings \cite{andersson87a,andersson87b,sjoestrand94a}. 
The leading hadrons of the fragmenting strings contain the valence-quarks of the original excited hadron. In UrQMD they are allowed to interact even during their formation time, with a reduced cross-section defined by the additive quark model, thus accounting for the original valence quarks contained in that hadron \cite{UrQMD}. 

Within the model, antibaryons are produced through string fragmentation. The Field$-$Feynman fragmentation mechanism \cite{Feynman}, which allows the independent string decay from both ends of the string, is used in the UrQMD model~\cite{UrQMD}. The string break-up is treated iteratively: String $\rightarrow$ hadron + smaller string. The conservation laws are fulfilled. The essential part of this mechanism is the fragmentation function which yields the probability distribution 
%$p(z^{\pm}_{\rm fraction}, m_t)$. 
$p(z^{\pm}_{\rm fraction})$. 
This function regulates the fraction of energy and momentum given to the produced hadron in the stochastic fragmentation of the color string. For newly produced particles the Field-Feynman function~\cite{Feynman}:
\begin{equation}
p(z^{\pm}_{\rm fraction})={\rm constant}\times(1-z^{\pm}_{\rm fraction})^2\,,
\end{equation}
is used. $P(z)$ drops rapidly with increasing $z$. %(Fig.~\ref{fig21}). 
Therefore, the longitudinal momenta of the produced antibaryons are small; they are mostly produced in the central rapidity region with high baryon densities~\cite{bleicherReview}. 

%Particle production from string fragmentation basically follows the Boltzmann statistical law of $\exp(-m/T)$ where $m$ is the particle mass and $T\approx 140$~MeV/c which is related to the string tension.
At AGS energies, antibaryon production is very rare. In order to increase statistics, we modified the string fragmentation routine in such a way that every string fragmenting process is repeated to a maximum of 1000 times or until at least one antibaryon or antibaryon resonance is produced. This drastically increased the absolute abundance of antibaryons, but does not alter the relative abundance among antibaryons.
%As seen from the figure, less than 1 in a 1000 survives the detector. To increase the statistics, we insist that UrQMD always produces an antibaryon from each string. This requirement does not alter the \lp\ because the relative abundance of $\pbar$ and higher mass baryons is fixed on average in each string fragmentation.

\begin{figure}
\centerline{\includegraphics[width=0.45\textwidth]{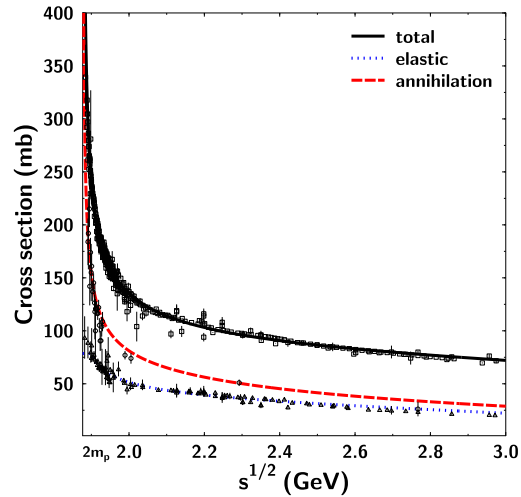}}
\caption{(Color online) Parameterization (dashed curve) to the measured $\pbarp$ annihilation cross-section (open circles) along with those of elastic and total cross-sections. The parametrization is used in the UrQMD model. The $\lbarp$ annihilation cross-section used in the model is deduced from that of $\pbarp$ by the Additive Quark Model.}
\label{fig:xsec}
\end{figure}

Once produced, an antibaryon may or may not annihilate with baryons in the collision zone. The $\pbarp$ annihilation cross-section is well measured. Figure~\ref{fig:xsec} depicts the annihilation cross-sections together with the elastic and total cross-sections~\cite{xsec}. In this study, the $\bar{p}p$ annihilation cross-section is given by 
\begin{equation}
\sigma_{\pbarp}^{\rm annih}(\sqrt{s})=1.2\frac{\sigma_{\pbarp}^{\rm total}(\sqrt{s})}{\sqrt{s}}\,,
\end{equation}
where $\sqrt{s}$ is in~GeV. The total $\bar{p}p$ cross-section is taken from the CERN-HERA parametrization~\cite{xsec} (shown in Fig.~\ref{fig:xsec}). The other antibaryon-baryon annihilation cross-sections are, however, not well measured. UrQMD applies a correction factor, given by the Additive Quark Model for these annihilation cross-sections~\cite{bleicherReview}:
\begin{equation}
\frac{\sigma_{\bar{B}B}(\sqrt{s})}{\sigma_{\pbarp}(\sqrt{s})}=\left(1-0.4\frac{s_{\bar{B}}}{3}\right)\left(1-0.4\frac{s_B}{3}\right)\,.
\label{eq:aqm}
\end{equation}
Here $s_{\bar{B}}$ is the strangeness number of the antibaryon and the baryon, respectively. There is a reduction of 13\% due to each strange or anti-strange quark. For instance, the $\lbarp$ annihilation cross-section is 
\begin{equation}
\sigma_{\lbarp}(\sqrt{s})=0.87 \sigma_{\pbarp}(\sqrt{s})\,.
\label{eq:xsec}
\end{equation}
Note that the relationship in Eq.~(\ref{eq:xsec}) is for the same c.m.s.~energy $\sqrt{s}$ of the $\lbarp$ and $\pbarp$ systems. Generally, the $\lbarp$ c.m.s.~energy is larger than that of $\pbarp$ in heavy-ion collisions, so the reduction factor is lower than 0.87. The average c.m.s.~energy square of a pair of particles in a chaotic system is approximately $s\approx m_1^2+m_2^2+2E_1E_2$, where $m_i$ and $E_i$ are the rest mass and total energy of the particles. The transverse distributions of $\pbar$ and $\lbar$ have been measured at rapidity about 1.2~\cite{E917,E866pbar}. Taken the mid-rapidity transverse distributions to be similar, the average energies of the $\pbar$ and $\lbar$ at mid-rapidity are roughly 1.18 and 1.42~GeV, respectively. The c.m.s.~energies of $\pbarp$ and $\lbarp$ pairs are therefore 2.13 and 2.34~GeV, respectively. This difference in $\sqrt{s}$ introduces an additional reduction of the $\lbarp$ annihilation cross-section by approximately 23\% as opposed to the $\pbarp$ annihilation cross-section. Therefore, the effective relationship between the $\pbarp$ and $\lbarp$ annihilation cross-sections at the AGS is 
\begin{equation}
%\sigma_{\lbarp}(\sqrt{s})\approx0.67\sigma_{\pbarp}(\sqrt{s})\,.
\langle\sigma_{\lbarp}\rangle\approx0.67\langle\sigma_{\pbarp}\rangle\,.
\label{eq:xsec2}
\end{equation}

\begin{figure}
%\centerline{\includegraphics[width=0.45\textwidth]{UrQMD-1.1/ann.eps}}
\centerline{\includegraphics[width=0.45\textwidth]{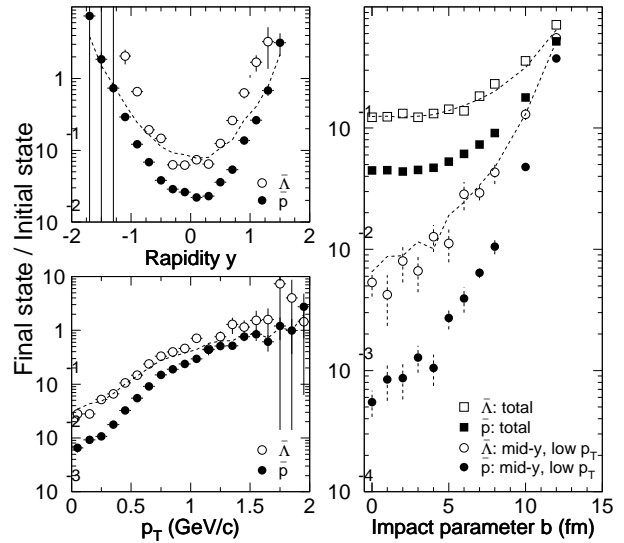}}
\caption{The ratio of number of $\lbar$'s (open symbols) and $\pbar$'s (solid symbols) at final freeze-out over that at the initial string fragmentation stage, calculated by UrQMD for central ($b<1$ fm) Au+Au collisions at $\snn=4.9$~GeV, as a function of (a) rapidity but integrated over $p_T$, and (b) $p_T$ but integrated over rapidity, and for all (minimum bias) collisions, as a function of (c) impact parameter $b$ for the mid-rapidity ($|y|<0.4$) and low $p_T$ ($p_T<0.3$ GeV/$c$) region (circles) as well as for integrated over whole phase-space (squares). At AGS energies, antibaryons can be produced only at the very early time from string fragmentation, and are then annihilated by baryons at later times. However, the ratio can be larger than one because the final freeze-out rapidity and/or $p_T$ distribution can be broader than the initial distributions. With this caution, the ratio may be viewed as the ``survival probability'' of $\lbar$ and $\pbar$ through the nuclear matter created in the collisions. The dashed curves indicate a simple optical model prediction of the $\lbar$ survival probability using the $\pbar$ result and the different annihilation cross-sections of $\lbar$ and $\pbar$ with nucleons. See the text.}
\label{fig:ann}
\end{figure}

Figure~\ref{fig:ann} gives an idea about the magnitude of the absorption effect in Au+Au collisions by plotting the ratio of freeze-out $\lbar$ (or $\pbar$) over that at the initial production stage. The ratio can be viewed as the ``survival probability'' of $\lbar$ (or $\pbar$) from initial production to the final freeze-out. The ratio is higher than one in forward and backward rapidities because the final $dN/dy$ distributions can be broader than the initial ones. The left upper plot shows the ``survival probability'' as a function of rapidity, and the left lower plot shows that as a function of $p_T$ for central Au+Au collisions. It is clearly seen that the largest absorption is in the mid-rapidity and low $p_T$ region. The right plot shows the ``survival probabilities'' of mid-rapidity ($|y|<0.4$) and low $p_T$ ($p_T<0.3$~GeV/$c$) $\lbar$'s and $\pbar$'s in solid symbols and integrated over whole phase-space in open symbols, as a function of impact parameter ($b$) in Au+Au collisions. As seen from the plot, about 99.9\% and 99\% of the mid-rapidity and low $p_T$ $\pbar$'s and $\lbar$'s produced in central Au+Au collisions are annihilated. In other words, only 1 out of 1000 $\pbar$'s and 1 out of 100 $\lbar$'s in this kinematic region survive to freeze-out.

On the other hand, if all the $\lbar$'s and $\pbar$'s are counted, then the ``survival probabilities'' are much higher, and are roughly constant over a wide range of impact parameters in central collisions. However, this has implications on the interpretations of the measured absolute $\pbar$ yields. The measured yields (within a fixed rapidity window) has a less than linear increase with the total number of participants in Au+Au collisions~\cite{E866pbar}. The power factor of the increase is 0.74. If we take into account the absorption effect shown in the open symbols in Fig.~\ref{fig:ann}(c), then the restored initially produced $\pbar$'s would have a stronger than linear increase with the total number of participants. The power factor would be about 1.46. 

\begin{figure}
%\centerline{\includegraphics[width=0.5\textwidth]{UrQMD-1.1/fo_b.eps}}
\centerline{\includegraphics[width=0.5\textwidth]{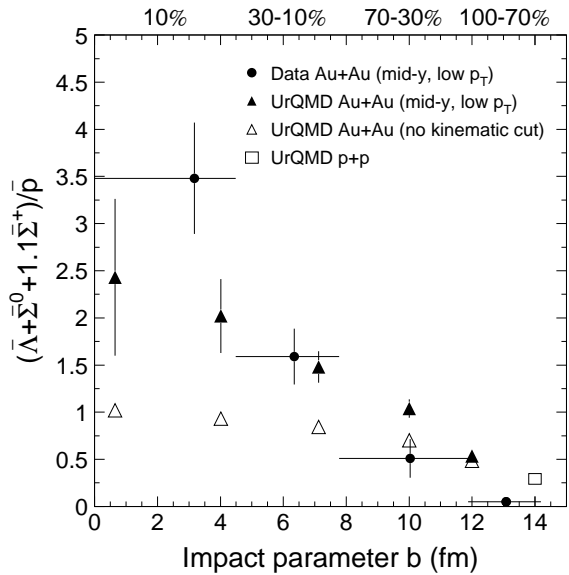}}
\caption{UrQMD calculation of the freeze-out \LP\ ratio at mid-rapidity ($|y|<0.4)$ and low $p_T$ ($p_T<0.3$ GeV/$c$) as a function of impact parameter $b$ (solid triangles) in Au+Au collisions at $\snn=4.9$~GeV, compared to experimental results in similar kinematic region (solid circles). Note that UrQMD can reasonably reproduce the data. The UrQMD total yield \LP\ ratio (no kinematic cut) is shown as open triangles. The UrQMD total yield \LP\ ratio in $p+p$ collisions at $\sqrt{s}=4.9$~GeV is indicated by the box drawn at $b=14$~fm.}
\label{fig:freezeout}
\end{figure}

Now back to Fig.~\ref{fig:ann}. In the simple picture of a sphere of baryons with a uniform density $\rho$ and radius $R$, the survival probability of an antibaryon produced at the center is
\begin{equation}
P_{\rm surv}=\exp(-\sigma\rho L)\,,
\label{eq:surv}
\end{equation}
where $\sigma$ is the annihilation cross-section. 
%In order to get an intuitive idea, one may recall Eq.~(\ref{eq:xsec}) and scale the $\pbar$ results by a power factor smaller than 0.87 to obtain the $\lbar$ survival probability. Adding in an {\it ad hoc} power factor of 2/3, 
By using the factor 0.67 from Eq.~(\ref{eq:xsec2}) as our power factor, we have
\begin{equation}
P_{\lbar,\rm surv}=P_{\pbar,\rm surv}^{0.67}\,,
\label{eq:surv2}
\end{equation}
and obtain the dashed curves in Fig.~\ref{fig:ann} given the $\pbar$ data points. We find good agreement with the calculated $\lbar$ survival probability except for the forward and backward rapidity regions.

Figure~\ref{fig:freezeout} shows the freeze-out ratio of \LP\ at mid-rapidity ($|y|<0.4$) and low $p_T$ ($p_T<0.3$~GeV/$c$) in Au+Au collisions at AGS energy as a function of the collision impact parameter $b$. The experimentally deduced ratio is reproduced from Ref.~\cite{E864PRC} with $b$ values obtained from the centrality bins. The UrQMD ratio is in a good agreement with the data. The \LP\ ratio of the total particle yields are also shown. It is clear that the large ratio in the mid-rapidity and low $p_T$ region in central collisions is largely due to the kinematic cut. Note that the total yield \LP\ ratio in $p+p$ interactions as calculated by UrQMD is in a good agreement with the trend from the heavy-ion results.

%fqw
As UrQMD successfully describes the data, it is interesting to examine the ratio at the initial production stage. Since high-mass antibaryon resonances are present at the initial stage, it is important to correct for the feed-down contributions of these resonances to $\pbar$'s, $\lbar$'s, and $\bar{\Sigma}$'s. 
%The properly feeddown corrected \LP\ ratio is shown in Fig.~\ref{fig:initial}. The ratio is roughly independent of centrality. The \LP\ ratio in isospin average nucleon-nucleon (N+N) collisions calculated by UrQMD is indicated in the upper box drown at $b=14$~fm. This value seems not to follow extrapolations from the heavy-ion results. However, it may be understandable because there may be larger effect of feeddowns from high mass resonance in \LP\ ratio in heavy-ion collisions than in N+N collisions.
We find that the initial ratios of \LP\ $\approx0.46$ and \lp\ $\approx0.14$ are below unity and are roughly independent of centrality. The initial \lp\ ratio is consistent with that calculated in isospin averaged nucleon-nucleon collisions. The centrality independence can be readily understood because the string fragmentation function does not know about the centrality of the collision and there are no other antibaryon production mechanisms in UrQMD that are centrality-dependent (e.g.~due to QGP production).
%
%fqw
%The most interesting question to ask is what the pure \lp\ ratio (without feeddowns) is because a significantly larger value of this ratio in heavy-ion collisions than in N+N collisions may signal formation of the QGP. This ratio calculated by UrQMD is shown in Fig.~\ref{fig:initial} as a function of impact parameter in Au+Au collisions. The ratio is independent of centrality. This is so because string fragmentation function does not know about the centrality of the collisions, and there is no QGP in UrQMD. The \lp\ ratio in N+N collisions as calculated by UrQMD is shown in the lower box drawn at $b=14$~fm. This value is similar to the experimental measurement of about 0.2 in $p+p$ collisions~\cite{Blobel,Rossi}. The heavy-ion \lp\ ratios agree reasonably well with the N+N value. 

As UrQMD well reproduces the freeze-out values for the \LP\ ratio, it is reasonable to suggest that the data indicates no large ratio of \lp\ at the initial stage, and therefore does not imply QGP formation.

The \Lp\ ratio has been also measured in heavy-ion collisions at the SPS. %Data on the \Lp\ ratio are eminent from collisions at RHIC. 
The data show a decrease of the ratio from central to peripheral colllisions~\cite{NA49lbar}. The decrease was also observed in UrQMD where the $\bar{\Lambda}/\bar{p}$ ratio drops rapidly with increasing $b$ from 1.3 to 0.5 \cite{Bleicher:1999xi}. This suggests an interplay between particle production and subsequent annihilation also at the SPS energies. In peripheral (large $b$) collisions the $\bar{\Lambda}$ production is basically the same as in $p+p$ reactions. $\bar{\Lambda}$'s and $\bar{p}$'s are produced via the fragmentation of color flux tubes (strings). The production of (anti-)strange quarks in the color field is suppressed due to the mass difference between strange and up and down quarks. This results in a suppression of $\bar{\Lambda}$ over $\bar{p}$ by a factor of 2 ($\bar{\Lambda} /\bar{p}\approx 0.3$-0.5 in $p+p$). %Note that no additional cuts have been applied at SPS energies. This is consistent with experimental measurement~\cite{NA49lbar}.

The \Lp\ ratio in high energy heavy-ion collisions at RHIC was measured to be singinificantly smaller than that at the AGS and SPS energies~\cite{STAR_lbar,PHENIX_lbar}. This is an experimental demonstration of the importance of the net-baryon density on the \Lp\ ratio. At the top RHIC energy, the net-baryon density is small at mid-rapidity~\cite{RHIC_pbar,RHIC_netBaryon}, resulting in an insignificant effect on the \Lp\ ratio. The \Lp\ ratio measured at RHIC, therefore, reflects more truly the initial production stage. This also demonstrates that a mere large \Lp\ ratio is by no means a signature of the QGP formation, as the QGP formation is much more likely at RHIC than at the AGS. One has to take into account the effect of antibaryon absorption before connecting a large \Lp\ ratio to the QGP formation.

\begin{figure}
%\centerline{\includegraphics[width=0.5\textwidth]{UrQMD-1.1/initial.eps}}
%\centerline{\includegraphics[width=0.5\textwidth]{initial.eps}}
%\caption{The pure \lp\ ratio (without feeddowns) and the \LP\ ratio (with feeddowns) at the initial string fragmentation stage as calculated by UrQMD in Au+Au collisions at $\snn=4.9$~GeV as a function of impact parameter. The corresponding results in $p+p$ collisions at $\sqrt{s}=4.9$~GeV from UrQMD are indicated by the boxes at $b=14$~fm. Note that the ratios in Au+Au collisions is constant over centrality. The \lp\ ratio in Au+Au colllisions agrees with the $p+p$ result, as expected from string fragmentation to be the only production mechanism for antibaryons at the AGS. %UrQMD results have no kinematic cut (all y, all pt).}
%\label{fig:initial}
\centerline{\includegraphics[width=0.5\textwidth]{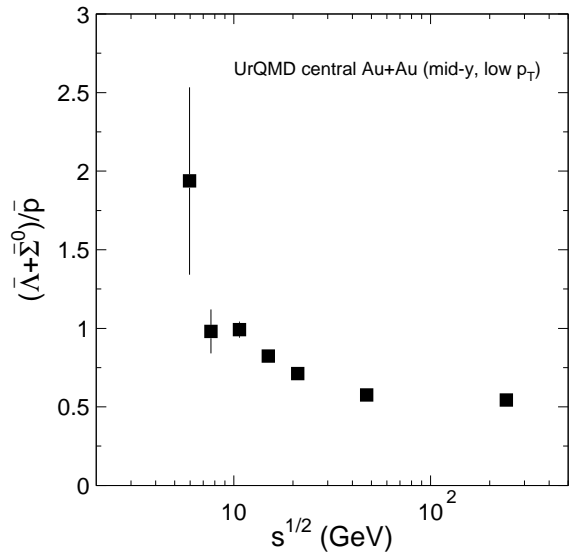}}
\caption{The excitation function of the \Lp\ ratio from UrQMD in the AGS, SPS, and RHIC energy range. The UrQMD calculations were done for $|y|<0.4$ and $p_T<0.3$~GeV/$c$ in central Au+Au collisions ($b<1$~fm).}
\label{fig:roots}
\end{figure}

In Fig.~\ref{fig:roots} we show the \Lp\ ratio from  UrQMD as a function of the collision energy $\snn$. Here we have kept the same parameters as for the previous AGS analysis for consistency, i.e.~a cut on mid-rapidity and low $p_T$ and the same centrality. The ratio is found to decrease with $\snn$ because of the decreasing net-baryon density with increasing collision energy.

In conclusion, a large \lp\ ratio is observed in central Au+Au collisions at the AGS. A strong increase of the \lp\ ratio from peripheral to central collisions is indicated from the experimental data. The hadronic transport model, UrQMD, can satisfactorily describe the large ratio in central collisions and the centrality dependence of the ratio. According to the model, the experimentally deduced large ratio of \LP\ in the mid-rapidity and low $p_T$ region is mainly due to the strong and different absorption of these antibaryons. 
%Therefore, the data do not exclusively imply a QGP formation.
%To fully understand the $\bar{\Lambda}/\bar{p}$ ratio in heavy-ion collisions, measurements of low energy $\bar{\Lambda}N$ annihilation cross-sections are needed.
The measured large \lp\ ratio in itself is by no means a signature of the QGP formation.
An increase of the \lp\ ratio from peripheral to central collisions was also observed at the SPS, but not as large as that at the AGS. The \lp\ ratio in central heavy-ion collisions was found to steadily decrease with increasing collision energy from the AGS, SPS, to RHIC. An excitation function measurement of the \lp\ ratio, especially the energy region from SPS to the top RHIC energy, will be valuable. This is presently being undertaken by the beam energy scan at RHIC.

%\acknowledgments{
This work is partially supported by the U.S. Department of Energy under Grant No.~DE-FG02-88ER40412 and the RFBR-CNRS Grants No.~08-02-92496-NTsNIL\_a and No.~10-02-93111-NTsNIL\_a. We thank Lawrence Berkeley National Laboratory where this work was initiated.

%%%%%%%%%%%%%%%%%%%%%%%%%%%%%%%%%%%%%%%%%%%%%%%%%%%%%%%%%%%%%%%%%%%%%%%%%%%

%%%%%%%%%%%%%%%%%%%%%%%%%%%%%%%%%%%%%%%%%%%%%%%%%%%%%%%%%%%%%%%%%%%%%%%%%%%

\end{document}